\documentclass[a4paper]{article}
\usepackage{amsmath,amssymb,theorem,fullpage}
\usepackage[shortcuts]{extdash}
\usepackage{hyperref}
\usepackage[nosort]{cite}

\let\frac\undefined

\allowdisplaybreaks

\numberwithin{equation}{section}

\def\Maketitle{{\def\newpage{}\maketitle}}

\def\eq#1$$#2$${\begin{equation#1}#2\end{equation#1}}
\long\def\subeq#1{\begin{subequations}#1\end{subequations}}
\def\Split$$#1$${\begin{split}#1\end{split}}
\def\Align#1$$#2$${\begin{align#1}#2\end{align#1}}
\def\AlignAt#1$$#2$${\begin{alignat}{#1}#2\end{alignat}}
\def\Aligned#1{\begin{aligned}#1\end{aligned}}
\def\Gather#1$$#2$${\begin{gather#1}#2\end{gather#1}}
\def\Gathered#1{\begin{gathered}#1\end{gathered}}
\def\Multline#1$$#2$${\begin{multline#1}#2\end{multline#1}}

\def\d{\partial}

\def\Im{\mathop{\rm Im}\nolimits}

\def\Res{\mathop{\rm Res}\limits}

\def\cA{{\mathcal A}}
\def\cF{{\mathcal F}}
\def\bcF{{\bar{\mathcal F}}}

\def\cO{{\mathcal O}}
\def\cR{{\mathcal R}}
\def\cS{{\mathcal S}}

\def\sh{\mathop{\rm sh}\nolimits}
\def\ch{\mathop{\rm ch}\nolimits}

\def\lcolon{\mathopen{\,:}}
\def\rcolon{\mathclose{:\,}}

\def\R{{\mathbb R}}
\def\Z{{\mathbb Z}}

\def\e{{\rm e}}
\def\i{{\rm i}}

\def\ba{{\bar a}}

\def\bh{{\bar h}}

\def\bm{{\bar m}}
\def\bpi{{\bar\pi}}

\def\lambdadag{\lambda^{\smash\dagger}}

\def\tdag{t^{\smash\dagger}}

\def\Rdag{R^{\smash\dagger}}

\makeatletter

\def\section{\@startsection{section}{1}{\z@}%
                                   {-3.5ex \@plus -1ex \@minus -.2ex}%
                                   {2.3ex \@plus.2ex}%
                                   {\normalfont\normalsize\bfseries}}
\def\subsection{\@startsection{subsection}{2}{\z@}%
                                     {-3.25ex\@plus -1ex \@minus -.2ex}%
                                     {1.5ex \@plus .2ex}%
                                     {\normalfont\normalsize\bfseries\itshape}}
\def\@seccntformat#1{\csname the#1\endcsname.~~}
\long\def\@makecaption#1#2{%
  \vskip\abovecaptionskip
  \sbox\@tempboxa{\small#1. #2}%
  \ifdim \wd\@tempboxa >0.9\hsize
  {\leftskip=0.05\hsize\rightskip=0.05\hsize\relax\small
    #1. #2\par}
  \else
    \global \@minipagefalse
    \hb@xt@\hsize{\hfil\box\@tempboxa\hfil}%
  \fi
  \vskip\belowcaptionskip}
\def\Appendix{\appendix
  \def\@seccntformat##1{Appendix~\csname the##1\endcsname.~~}}

\let\over\@@over
\let\atop\@@atop
\let\above\@@above
\let\overwithdelims\@@overwithdelims
\let\atopwithdelims\@@atopwithdelims
\let\abovewithdelims\@@abovewithdelims

\makeatother

\long\def\?#1{{\par\medskip\hrule\smallskip\noindent
{\bf What is missing:} #1\smallskip\hrule\medskip\par}}

\begin{document}


\title{Algebraic approach to form factors\\
in the complex sinh-Gordon theory}
\author{Michael Lashkevich and Yaroslav Pugai,\\[\medskipamount]
\parbox[t]{0.9\textwidth}{\normalsize\it\raggedright
Landau Institute for Theoretical Physics, 142432 Chernogolovka, Russia\medspace%
\footnote{Mailing address.}
\\
Moscow Institute of Physics and Technology, 141707 Dolgoprudny, Russia\\
Kharkevich Institute for Information Transmission Problems, 19 Bolshoy Karetny per., 127994 Moscow, Russia}
}
\date{}

\Maketitle

\begin{abstract}

We study form factors of the quantum complex sinh\-/Gordon theory in the algebraic approach. In the case of exponential fields the form factors can be obtained from the known form factors of the $Z_N$\-/symmetric Ising model. The algebraic construction also provides an Ansatz for form factors of descendant operators. We obtain generating functions of such form factors and establish their main properties: the cluster factorization and reflection equations.

\end{abstract}

\section{Introduction}
We study form factors of local and quasilocal operators in the two\-/dimensional complex sinh\-/Gordon model, which is a quantum version of the model introduced by Pohlmeyer\--Lund\--Regge~\cite{Pohlmeyer:1975nb,Lund:1976ze} for negated coupling constant. The action of the theory is given by
\eq$$
\cS[\chi,\bar\chi]=\int {d^2x\over4\pi}\,\left({\d_\mu\chi\,\d^\nu\bar\chi\over1+g_0\chi\bar\chi}-m_0^2\chi\bar\chi\right),
\label{CShG-action}
$$
where $\chi=\chi_1+\i\chi_2$, $\bar\chi=\chi_1-\i\chi_2$ is a complex scalar boson field. Initially, the model was introduced for negative values of $g_0$, where it has a rich spectrum of solitons and their bound states. We will assume that $g_0>0$, so that the spectrum consists of the only particle\-/antiparticle pair~\cite{Fateev:1995ht}. De Vega and Maillet~\cite{deVega:1981ka,deVega:1982sh} found that in the semiclassical limit the coupling constant is finally renormalized according to
\eq$$
g={g_0\over1+g_0},
\label{g-renorm}
$$
while the mass of the lightest particle, which corresponds to the only particle of the model we consider, is given by
\eq$$
m=m_0{\sin\pi g\over\pi g}.
\label{m-renorm}
$$
The exact quantum relations are unknown, but our results do not rely upon them, since we always assume $g$ be a constant in the exact $S$ matrix below and $m$ be the mass of the particle.

The model is integrable on the quantum level~\cite{Bonneau:1984pj}. It means that the $S$ matrix is factorizable and, hence, is uniquely defined as soon as the two-particle $S$ matrix is known. Denote the particle by $1$ and the antiparticle by~$\bar1$. According to Dorey and Hollowood~\cite{Dorey:1994mg} the exact two-particle $S$ matrix reads
\eq$$
S_{11}(\theta)=S_{\bar1\bar1}(\theta)=S_{1\bar1}(i\pi -\theta)
={\sinh\left(\theta/2 -\i\pi g\right)\over \sinh\left(\theta/2+\i\pi g\right)},
\label{S-matrix}
$$
where $\theta=\theta_1-\theta_2$ is the difference of rapidities $\theta_i$ of colliding particles, defined in the standard way in terms of momenta: $p_i=(m\ch\theta_i,m\sh\theta_i)$. In our case $g>0$ there are no poles on the physical sheet ($0\le\Im\theta\le\pi$), so that there are no bound states in the theory.

Form factors are matrix elements of quasilocal operators of a theory with respect to the eigenstates of the Hamiltonian. A full set of form factors completely defines an operator. Form factors make it possible to calculate large-distance asymptotics of correlation functions in massive models with, in principle, arbitrary precision. For integrable models on the plane form factors can be found exactly by solving a system of linear difference equations and analyticity conditions called form factor axioms~\cite{Smirnov:1992vz}, as soon as the spectrum of the model and the $S$ matrix are known. Every solution defines a quasilocal operator. There are many approaches to find solutions to the form factor axioms~\cite{Smirnov:1990vm,Lukyanov:1993pn,Babujian:2002fi}. In this note we show how to find form factors for the complex sinh\-/Gordon model following the algebraic approach proposed in~\cite{Feigin:2008hs,Alekseev:2009ik,Lashkevich:2013mca}

More precisely, let $|\theta_1\alpha_1\ldots\theta_N\alpha_N\rangle$ be the eigenstate defined as an in-state with $N$ particles of types $\alpha_1,\ldots,\alpha_N$ with rapidities $\theta_1>\ldots>\theta_N$. Let $\cO(x)$ be any quasilocal operator. Then the matrix elements
\eq$$
\langle{\rm vac}|\cO(0)|\theta_1\alpha_1\ldots\theta_N\alpha_N\rangle=F_\cO(\theta_1,\ldots,\theta_N)_{\alpha_1\ldots\alpha_N}
\label{FO-def}
$$
define a set of analytic functions $F_\cO$. All other matrix elements are expressed in terms of these analytic functions by means of the crossing symmetry.

To understand the operator contents of the theory, its dual description~\cite{Fateev:1995ht,Fateev:2001mj} is more useful. The dual action is formulated in terms of two neutral scalar fields $\vartheta(x),\varphi(x)$:
\eq$$
\cS_{\rm dual}[\vartheta,\varphi]=\int d^2x\,\left({(\d_\mu\vartheta)^2+(\d_\mu\varphi)^2\over8\pi}
+m\e^{\beta\varphi}\cos\alpha\vartheta-\lambda\e^{-2\beta\varphi}\right).
\label{dual-action}
$$
Here the coupling constants $\alpha$ and $\beta$ are not independent, and they are related to the coupling constant $g$ of the complex sinh-Gordon model:
\eq$$
\alpha^2=\beta^2+1={1\over2g}.
\label{alpha-beta-def}
$$
The constant $\lambda$ is of dimension $m^{2+4\beta^2}$ and by an appropriate shift of the field $\varphi$ can be adjusted in such a way that $m$ would coincide with the physical mass of the theory. Introduce the exponential operators
\eq$$
V_{abq}(x)
=\exp\left({\i a\over2\alpha}\vartheta(x)-{b\over2\beta}\varphi(x)-{\i q\over2\alpha}\tilde\vartheta(x)\right).
$$
Here $a,b$ are arbitrary real parameters, an integer $q$ is the charge of the operator, and $\tilde\vartheta(x)$ is the dual field defined, as usual, by the equation $\d^\mu\tilde\vartheta=\epsilon^{\mu\nu}\d_\nu\vartheta$. The fields $\chi,\bar\chi$ are given (up to a normalization constant) by
\eq$$
\chi(x)\sim V_{01\,1}(x),
\qquad
\bar\chi(x)\sim V_{01\,-1}(x),
\qquad
\chi\bar\chi(x)\sim V_{02\,0}(x).
\label{chi-barchi-V}
$$
Since the expressions above contain the dual field $\tilde\vartheta$, these operators are not mutually local. Two operators $\cO_1,\cO_2$ are mutually quasilocal, if they possess the following property. Consider any correlation function $\langle\cO_1(x_1)\cO_2(x_2)\cdots\rangle$. If we take $x_1$ and make a round trip in the counter\-/clockwise direction along a closed contour, which encircles $x_2$, we will just gain a phase factor $\e^{2\pi\i\gamma}$. The quantity $\gamma=\gamma(\cO_1,\cO_2)$ is called the mutual locality exponent of the operators. For the exponential operators $V_{abq}(x)$ the mutual locality exponents are
\eq$$
\gamma(V_{abq},V_{a'b'q'})=-g(aq'+a'q).
\label{VV-locality}
$$
In particular, the mutual locality with the bosons $\chi,\bar\chi$ are
\eq$$
\gamma(V_{abq},\chi)=-\gamma(V_{abq},\bar\chi)=-ga.
\label{Vchi-locality}
$$
It means that the operator $V_{abq}$ is local with respect to the basic bosons of the initial Lagrangian~(\ref{CShG-action}) for $ga\in\Z$ only.

Beside the exponential (or primary) operators $V_{abq}(x)$ the theory contains a set of descendant operators. Let us use the light-cone variables $x^\pm=x^1\pm x^0$ and the corresponding derivatives~$\d_\pm$. Then the operators of the form
\eq$$
\cO(x)=(\d_-^{k_1}\vartheta\cdots\d_-^{l_1}\varphi\cdots\d_+^{\bar k_1}\vartheta\cdots\d_+^{\bar l_1}\varphi\cdots)V_{abq}(x)
\label{descendant-def}
$$
are called level $(k,\bar k)$ descendant operators, where $k=\sum k_i+\sum l_i$, $\bar k=\sum\bar k_i+\sum\bar l_i$. The (Lorentz) spin $s$ of such operator and its (ultraviolet) scaling dimension $d$ read
\eq$$
s=-gaq+k-\bar k,
\qquad
d={a^2+q^2\over4\alpha^2}-{b^2\over4\beta^2}+k+\bar k.
\label{sd-exponetial}
$$
It must be noted that the spin is well\-/defined in the massive theory, while the scaling dimension is not, so that operators can be renormalized by admixture of operators of lower dimensions.

The fields $V_{abq}(x)$ and their descendants are not all independent. It can be found that they are subject to two constraints called the reflection relations:
\eq$$
V_{abq}(x)=R^{\rm L}(b)V_{a,2g^{-1}-2-b,q}(x)=R^{\rm SL}(a,b,q)V_{a,-2-b,q}(x).
\label{Vabq-reflection}
$$
Here $R^{\rm L}(b)$ is the Liouville reflection function~\cite{Zamolodchikov:1995aa} and $R^{\rm SL}(a,b,q)$ is the sine-Liouville reflection function~\cite{Baseilhac:1998eq,Fateev:2001mj}. Since we will not discuss here the overall normalization of operators, we are not interested in their exact form. In the case of descendant operators the reflection functions become matrices~\cite{Zamolodchikov:1995aa,Fateev:1998xb}. Later we derive these properties for operators defined by their form factors.

In the region $g<0$ (the complex \emph{sine}-Gordon model) the scattering theory based on (\ref{S-matrix}) will contain bound states. For integer values of $g^{-1}=-N$ it coincides (up to a reduction) with that of the $\Z_N$ Ising model. The last is a perturbation of the well-known $\Z_N$ parafermion models~\cite{Fateev:1985mm,Zamolodchikov:1986gh}, which contains the primary operators $\Phi^\kappa_{m\bm}(x)$, $\kappa=0,1,\ldots,N$, $m-\kappa,\bm-\kappa=0\bmod2$, and their $W$\-/algebraic descendants. It can be shown that $V_{abq}\sim\Phi^\kappa_{m\bm}$, where the parameters $a,b,q$ are related to $\kappa,m,\bm$ according to
\eq$$
a={\bm-m\over2},
\qquad
b=\kappa,
\qquad
q={m+\bm\over2}.
\label{abq-kmm}
$$
This identification~\cite{Fukuda:2001jd} follows from the fact that the sine\-/Liouville model, i.e.\ the model described by the action (\ref{dual-action}) with $\lambda=0$, is nothing but the $SL(2,\R)/U(1)$ coset conformal field theory, which is a noncompact (or negative real $N$) counterpart of the $\Z_N$ parafermion conformal models.

Since the form factors of all primary operators and some descendant operators in the $\Z_N$ Ising model are known explicitly~\cite{Jimbo:2000ff,Fateev:2006js,Fateev:2009kp,Babujian:2003za}, we may extend the corresponding expressions to the region $g>0$. Moreover, our algebraic approach makes it possible to systematically study descendant operators of arbitrary levels.

In Section~\ref{sec-primary} we introduce the algebraic construction for primary operators. Then, in Section~\ref{sec-recurrent} we obtain recurrent relations and establish the main properties of form factors of primary operators. In Section~\ref{sec-descendant} we define descendant operators in terms of their form factors and prove their cluster factorization property. In Section~\ref{sec-genfunc} we show that the form factors of primary operators can be considered as generating functions for form factors of descendant operators for generic values of $a,b$, and derive the reflection relations for descendant operators.

\section{Form factors of primary fields}
\label{sec-primary}

We are searching form factors as solutions to the equations called the form factor axioms~\cite{Smirnov:1992vz}. Remind their formulation for our case. Let $s_\cO$ be the Lorentz spin of the operator $\cO(x)$. Denote $\chi_1=\chi$, $\chi_{\bar1}=\bar\chi$. Then
\subeq{\label{ffax}
\Align$$
1.
&\ F_\cO(\theta_1+\vartheta,\ldots,\theta_N+\vartheta)_{\alpha_1\ldots\alpha_N}
=\e^{s_\cO\vartheta}F_\cO(\theta_1,\ldots,\theta_N)_{\alpha_1\ldots\alpha_N},
\label{ffax-spin}
\\
2.
&\ F_\cO(\theta_1,\theta_2,\ldots,\theta_N)_{\alpha_1\alpha_2\ldots\alpha_N}
=\e^{2\pi\i\gamma(\cO,\chi_{\alpha_1})}F(\theta_2,\ldots,\theta_N,\theta_1-2\pi\i)_{\alpha_2\ldots\alpha_N\alpha_1},
\label{ffax-cyclic}
\\
3.
&\ F_\cO(\ldots,\theta_i,\theta_{i+1},\dots)_{\ldots\alpha_i\alpha_{i+1}\ldots}
=S_{\alpha_i\alpha_{i+1}}(\theta_i-\theta_{i+1})F_\cO(\ldots,\theta_{i+1},\theta_i,\ldots),
\label{ffax-commut}
\\
4.
&\ \Res_{\vartheta'=\vartheta+\i\pi}F_\cO(\vartheta',\vartheta,\theta_1,\ldots,\theta_N)_{\bar11\alpha_1\ldots\alpha_N}
\notag
\\
&\ \qquad
=-\i\left(1-\e^{2\pi\i\gamma(\cO,\chi)}\prod^N_{i=1}S_{1\alpha_i}(\vartheta-\theta_i)\right)
F_\cO(\theta_1,\ldots,\theta_N)_{\alpha_1\ldots\alpha_N}.
\label{ffax-kinematic}
$$}%
The last equation provides the residue of the only singularity (kinematic pole) of form factors on the physical sheet.

Our first aim is to find form factors of exponential operators $V_{abq}(x)$. We obtain them by extending expressions for form factors of the $\Z_N$ Ising models known from~\cite{Jimbo:2000ff,Fateev:2006js}. In contrast to previous works, we use a new algebraic construction, which admits a straightforward generalization to descendant operators (see Section~\ref{sec-descendant}).

Our construction is based on the Heisenberg algebra generated by the elements $\alpha^\pm_k$, $\beta^\pm_k$ ($k\in\Z$, $k\ne0$), $\gamma^\pm$ with the only nonzero commutation relations
\eq$$
[\alpha^\varepsilon_k,\alpha^{\varepsilon'}_l]=kA^{(1)}_{\varepsilon k}\delta_{\varepsilon+\varepsilon',0}\delta_{k+l,0},
\qquad
[\beta^\varepsilon_k,\beta^{\varepsilon'}_l]=kA^{(2)}_{\varepsilon k}\delta_{\varepsilon+\varepsilon',0}\delta_{k+l,0},
\qquad
[\gamma^-,\gamma^+]=(\log\omega)^{-1},
\label{ddcommut}
$$
where
\eq$$
A^{(i)}_k={1\over2}(\omega^{k/2}-\omega^{-k/2})(\omega^{k/2}+(-1)^{k+i}\omega^{-k/2}).
\label{ABdef}
$$
In what follows the parameter $\omega$ will be related to the coupling constant of the model:
\eq$$
\omega=\e^{-2\pi\i g}.
\label{omega-def}
$$
Add two central elements $P_+$, $P_-$ and define the vacuum vectors $|1\rangle_p$, ${}_p\langle1|$ with $p=(p_+,p_-)$ by the conditions
\eq$$
\Gathered{
\alpha^\varepsilon_k|1\rangle_p=\beta^\varepsilon_k|1\rangle_p=0\quad(k>0),
\qquad
P_\pm|1\rangle_p=p_\pm|1\rangle_p,
\\
{}_p\langle1|\alpha^\varepsilon_{-k}={}_p\langle1|\beta^\varepsilon_{-k}=0\quad(k>0),
\qquad
{}_p\langle1|P_\pm={}_p\langle1|p_\pm,
\\
{}_p\langle1|\e^{m\gamma^-+(n-m)\gamma^+}|1\rangle_p=1.
}\label{1vacsdef}
$$
In fact, the vacuum vectors depend on the value of~$n$ in the last line, but we may ignore this subtlety. The Fock spaces over bra and ket vacuums will be respectively denoted as $\cF_p$ and $\bcF_p$.

It is convenient to introduce the normal product $\lcolon\cdots\rcolon$ with respect to these vacuum states such that $\lcolon \alpha^\varepsilon_k\alpha^{\varepsilon'}_{-k}\rcolon=\alpha^{\varepsilon'}_{-k}\alpha^\varepsilon_k$, $\lcolon\beta^\varepsilon_k\beta^{\varepsilon'}_{-k}\rcolon=\beta^{\varepsilon'}_{-k}\beta^\varepsilon_k$ ($k>0$). As for zero mode operators, we assume
$$
\lcolon\prod_i\e^{m_i\gamma^-+n_i\gamma^+}\rcolon=\e^{\gamma^-\sum_im_i+\gamma^+\sum_in_i}.
$$
Define the vertex operators
\eq$$
\lambda_\varepsilon(z)=\omega^{\gamma^\varepsilon}\exp\sum_{k\ne0}{\alpha^\varepsilon_k+\beta^\varepsilon_k\over k}z^{-k},
\qquad
\lambdadag_\varepsilon(z)=\omega^{-\gamma^\varepsilon}\exp\sum_{k\ne0}{\alpha^\varepsilon_k-\beta^\varepsilon_k\over k}z^{-k}.
\label{lambdadef}
$$
Products of vertex operators are given by
\eq$$
\Aligned{
\lambda_\varepsilon(z')\lambda_\varepsilon(z)
&=\lcolon\lambda_\varepsilon(z')\lambda_\varepsilon(z)\rcolon,
&\lambda_\varepsilon(z')\lambda_{-\varepsilon}(z)
&=f_\varepsilon\left(z\over z'\right)\lcolon\lambda_\varepsilon(z')\lambda_{-\varepsilon}(z)\rcolon,
\\
\lambdadag_\varepsilon(z')\lambdadag_\varepsilon(z)
&=\lcolon\lambdadag_\varepsilon(z')\lambdadag_\varepsilon(z)\rcolon,
&\lambdadag_\varepsilon(z')\lambdadag_{-\varepsilon}(z)
&=f_\varepsilon\left(z\over z'\right)\lcolon\lambdadag_\varepsilon(z')\lambdadag_{-\varepsilon}(z)\rcolon,
\\
\lambda_\varepsilon(z')\lambdadag_{-\varepsilon}(z)
&=g_\varepsilon\left(z\over z'\right)\lcolon\lambda_\varepsilon(z')\lambdadag_{-\varepsilon}(z)\rcolon,
&\lambda_\varepsilon(z')\lambdadag_\varepsilon(z)
&=\lcolon\lambda_\varepsilon(z')\lambdadag_\varepsilon(z)\rcolon,
\\
\lambdadag_\varepsilon(z')\lambda_{-\varepsilon}(z)
&=g_\varepsilon\left(z\over z'\right)\lcolon\lambdadag_\varepsilon(z')\lambda_{-\varepsilon}(z)\rcolon,
&\lambda_\varepsilon(z')\lambdadag_\varepsilon(z)
&=\lcolon\lambda_\varepsilon(z')\lambdadag_\varepsilon(z)\rcolon,
}\label{lambdalambdaprod}
$$
where $\varepsilon=\pm$ and
\eq$$
f_\varepsilon(z)={\omega^{\varepsilon/2}z-\omega^{-\varepsilon/2}\over z-1},
\qquad
g_\varepsilon(z)={\omega^{-\varepsilon/2}z+\omega^{\varepsilon/2}\over z+1}.
\label{fgdef}
$$

It can be seen that the vertex operators commute everywhere except the poles of their products. Equations (\ref{lambdalambdaprod}) provide a simple rule to calculate vacuum expectation values of products of vertex operators:
\Multline*$$
\langle\lambda_{\varepsilon_1}(x_1)\cdots\lambda_{\varepsilon_K}(x_K)
\lambdadag_{\varepsilon'_1}(y_1)\cdots\lambdadag_{\varepsilon'_L}(y_L)\rangle
\\*
=\prod^K_{i<j}\langle\lambda_{\varepsilon_i}(x_i)\lambda_{\varepsilon_j}(x_j)\rangle
\prod^L_{i<j}\langle\lambdadag_{\varepsilon'_i}(y_i)\lambdadag_{\varepsilon'_j}(y_j)\rangle
\prod^K_{i=1}\prod^L_{j=1}\langle\lambda_{\varepsilon_i}(x_i)\lambdadag_{\varepsilon'_j}(y_j)\rangle,
$$
where $\langle\cdots\rangle$ denote an average with respect to any vacuum $|1\rangle_p$.

Now define the currents
\eq$$ 
\Aligned{
t(z)
&=\omega^{P_+/4}\lambda_+(z)+\omega^{-P_+/4}\lambda_-(z),
\\
\tdag(z)
&=\omega^{P_-/4}\lambdadag_+(z)+\omega^{-P_-/4}\lambdadag_-(z).
}\label{ttdag-def}
$$
Below we will need the following property. The products $t(z')t(z)$ and $\tdag(z')\tdag(z)$ are regular for any values of $z',z$, while $t(z')\tdag(z)$ has the only pole at $z'=-z$:
\eq$$
t(z')\tdag(z)={\i Cz\over z'+z}\left(\omega^{P_+-P_-\over4}s_+(z)-\omega^{P_--P_+\over4}s_-(z)\right)+O(1)
\quad\text{as $z'\to-z$,}
\label{ttdag-pole}
$$
where $C=\i(\omega^{1/2}-\omega^{-1/2})=2\sin\pi g$ and
\eq$$
s_\varepsilon(z)=\lcolon\lambda_\varepsilon(-z)\lambdadag_{-\varepsilon}(z)\rcolon.
\label{s-def}
$$

Let $X=(x_1,\ldots,x_K)$, $Y=(y_1,\ldots,y_L)$. Define the functions
\eq$$
J_{KL,p}(X|Y)={}_p\langle1|
t(x_1)\cdots t(x_K)\tdag(y_1)\cdots\tdag(y_L)|1\rangle_p
\label{Jdef}
$$
and
\Multline$$
\cR_p(\theta_1,\ldots,\theta_K|\theta'_1,\ldots,\theta'_L)
=C^{-{K+L\over2}}\e^{{g\over2}(p_+-p_-)\left(\sum^K_{i=1}\theta_i-\sum^L_{j=1}\theta'_j\right)}
\\*
\times
\prod^K_{i<j}R(\theta_i-\theta_j)\prod^L_{i<j}R(\theta'_i-\theta'_j)
\prod^K_{i=1}\prod^L_{j=1}\Rdag(\theta_i-\theta'_j)
\label{cR-def}
$$
with
\eq$$
\Gathered{
R(\theta)={\sh{\theta\over2}\over\sh\left({\theta\over2}+\i\pi g\right)}\bar R(\theta),
\qquad
\Rdag(\theta)=\bar R^{-1}(\theta-\i\pi),
\\
\bar R(\theta)
=\prod^\infty_{n=1}{\Gamma\left({\i\theta\over2\pi}+g+n\right)\Gamma\left(-{\i\theta\over2\pi}+g+n\right)\Gamma^2(n-g)
  \over\Gamma\left({\i\theta\over2\pi}-g+n\right)\Gamma\left(-{\i\theta\over2\pi}-g+n\right)\Gamma^2(n+g)},
}\label{RRdag-def}
$$
It can be checked that the functions
\eq$$
F_p(\theta_1,\ldots,\theta_K,\theta'_1,\ldots,\theta'_L)
_{\underbrace{\scriptstyle1\ldots1}_K\underbrace{\scriptstyle\bar1\ldots\bar1}_L}
=J_{KL,p}(\e^{\theta_1},\ldots,\e^{\theta_K}|\e^{\theta'_1},\ldots,\e^{\theta'_L})
\cR_p(\theta_1,\ldots,\theta_K|\theta'_1,\ldots,\theta'_L)
\label{fp-def}
$$
satisfy the form factors axioms~(\ref{ffax}) and, hence, define some operators $\Phi_{pq}(x)$ with $q=L-K$. Namely, equation (\ref{ffax-spin}) is evident. Equations (\ref{ffax-cyclic}), (\ref{ffax-commut}) are satisfied due to the fact that $J_{KL,p}(X|Y)$ is symmetric in $x_i$ and in $y_j$ variables, while the function $R(\theta)$ satisfies the equations
$$
R(2\pi\i-\theta)=R(\theta),\qquad R(\theta)=S_{11}(\theta)R(-\theta).
$$
The kinematic pole condition (\ref{ffax-kinematic}) follows from equation~(\ref{ttdag-pole}). It can be checked directly along the guidelines of~\cite{Feigin:2008hs}. The derivation is based on the property: $f_+(\e^\theta)/f_-(\e^\theta)=S_{11}(\theta)$, $g_+(\e^\theta)/g_-(\e^\theta)=S_{1\bar1}(\theta)$.

For negative integer $g^{-1}$ these functions coincide with form factors of primary operators $\Phi^\kappa_{m\bm}$ of the $\Z_N$ Ising models for $p_\pm=1+N+\kappa\pm{1\over2}(m-\bm)$~\cite{Jimbo:2000ff,Fateev:2006js}. As it was discussed in Introduction, for positive $g$ these operators are identified with exponential operators in the complex sinh-Gordon model. Hence, the operators $\Phi_{pq}(x)$ coincide, up to a normalization factor, with $V_{abq}(x)$, if we set
\eq$$
p_\pm=1-g^{-1}+b\mp a.
\label{ab-pp}
$$
It is easy to see that the exponential factor in (\ref{cR-def}) just corresponds to the locality exponents~(\ref{Vchi-locality}).

There are evident periodicity properties of the functions $J_{KL,p}$:
\eq$$
J_{KL,(p_+,p_-)}(X|Y)=(-)^KJ_{KL,(p_++2g^{-1},p_-)}(X|Y)=(-)^LJ_{KL,(p_+,p_-+2g^{-1})}(X|Y),
\label{Jperiodicity}
$$
which lead to the corresponding relations for form factors.

\section{Recursion relations}
\label{sec-recurrent}

Consider the function $J_{K+1\>L,p}(x,X|Y)$ as a function of the variable~$x$. First, consider its asymptotics as~$x\to\infty$:
$$
J_{K+1\>L,p}(x,X|Y)\to C_{KL,p_+}J_{KL,p+\sigma}(X|Y)
$$
with the shift $\sigma=(1,-1)$ and the constant factor
\eq$$
C_{KL,z}=2\cos{\pi g\over2}(z-K+L).
\label{Cdef}
$$
The only singularities of this function are simple poles at the points $x=-y_i$, $i=1,\ldots,L$. Their residues are easily computable by using (\ref{ttdag-pole}) and~(\ref{s-def}):
$$
\Res_{x=-y_i}J_{K+1\>L,p}(x,X|Y)
=y_iR_{p,i}(X|Y)J_{K\>L-1,p}(X|\hat Y_i).
$$
Here $\hat Y_i=Y\setminus\{y_i\}$ and
\Multline$$
R_{p,i}(X|Y)
=2\i\sin\pi g\cdot\biggl(\omega^{p_+-p_-\over4}
\prod^K_{j=1}f_+\left(-{x_j\over y_i}\right)
\prod^L_{j=1\ (j\ne i)}f_-\left(y_j\over y_i\right)
\\*
-\omega^{p_--p_+\over4}
\prod^K_{j=1}f_-\left(-{x_j\over y_i}\right)
\prod^L_{j=1\ (j\ne i)}f_+\left(y_j\over y_i\right)
\biggr).
\label{Rpdef}
$$
These properties uniquely define the function $J_{K+1\>L,p}(x,X|Y)$ as an analytic function of~$x$:
\eq$$
J_{K+1\>L,p}(x,X|Y)
=C_{KL,p_+}J_{KL,p+\sigma}(X|Y)+\sum^L_{i=1}{y_iR_{p,i}(X|Y)\over x+y_i}J_{K\>L-1,p}(X|\hat Y_i).
\label{Jxrecinf}
$$
This relation expresses the function $J_{K+1\>L}$ in terms of $J_{KL}$ and $J_{K\>L-1}$ recursively. Similarly, we obtain a recursion relation in~$L$:
\eq$$
J_{K\>L+1,p}(X|y,Y)
=C_{LK,p_-}J_{KL,p-\sigma}(X|Y)
+\sum^L_{i=1}{x_iR_{p^*,i}(Y|X)\over y+x_i}J_{K-1\>L,p}(\hat X_i|Y),
\label{Jyrecinf}
$$
where $(p_+,p_-)^*=(p_-,p_+)$. Together with the initial condition
\eq$$
J_{00,p}(\varnothing|\varnothing)=1
\label{Jinit}
$$
the recursion relations (\ref{Jxrecinf}), (\ref{Jyrecinf}) uniquely define the functions~$J_{KL,p}(X|Y)$ and, hence, the form factors of primary operators.

The use of the recursion relations is that they make it possible to prove some important properties. First of all, we immediately find that the $J_{K0}$ and $J_{0L}$ functions are constant:
\eq$$
J_{K0,p}(X|\varnothing)=J_{K,p_+},
\qquad
J_{0L,p}(\varnothing|Y)=J_{L,p_-},
\qquad
J_{K,z}=\prod^K_{i=1}2\cos{\pi g\over2}(z+K+1-2i).
\label{J0}
$$
Next, the following symmetry properties can be easily obtained:
\Align$$
J_{KL,p}(X|Y)
&=J_{LK,p^*}(Y|X),
\label{ppexchange}
\\
J_{KL,p}(X|Y)
&=J_{KL,-p}(X^{-1}|Y^{-1}),
\label{ppneg}
\\
J_{KL,p}(X|Y)
&={J_{K-L,p_+}\over J_{K-L,-p_-}}J_{KL,-p^*}(X|Y).
\label{ppreflection}
$$
Equations (\ref{ppexchange}) and (\ref{ppneg}) are evident. The equation (\ref{ppreflection}), which corresponds to the reflection relations~(\ref{Vabq-reflection}), is a little less trivial. Literally, it corresponds to the Liouville reflection, but together with the periodicity properties~(\ref{Jperiodicity}) it provides the sine-Liouville reflection as well.

Let us prove (\ref{ppreflection}) from the recursion relations. For the sake of definiteness, let $K\ge L$. For $L=0$ this relation is an immediate consequence of~(\ref{J0}). We shall prove it by induction in~$L$. Assume it to hold for all $L\le L_0$. By applying it to the r.h.s.\ of (\ref{Jyrecinf}) for $L=L_0$, we obtain
\Multline*$$
J_{K\>L_0+1,p}(X|Y,y)
=C_{L_0K,p_-}{J_{K-L_0,p_+-1}\over J_{K-L_0,-p_--1}}J_{KL_0,-p^*-\sigma}(X|Y)
\\
+{J_{K-L_0-1,p_+}\over J_{K-L_0-1,-p_-}}
\sum^{L_0}_{i=1}{x_iR_{-p,i}(Y|X)\over y+x_i}J_{K-1\>L_0,-p^*}(\hat X_i,Y).
$$
We used the evident fact that $R_{p^*,i}(Y|X)=R_{-p,i}(Y|X)$. It is easy to check that
$$
C_{L_0K,p_-}{J_{K-L_0,p_+-1}\over J_{K-L_0,-p_--1}}
=C_{KL_0,p_+}{J_{K-L_0-1,p_+}\over J_{K-L_0-1,-p_-}}.
$$
By substituting it to the previous equation and applying (\ref{Jyrecinf}) again, we obtain (\ref{ppreflection}) for $L=L_0+1$.

\section{Form  factors of descendant fields}
\label{sec-descendant}

Let $\cA=\bigoplus^\infty_{k=0}\cA_k$ be a commutative graded algebra generated by the elements $a_{-k}$, $b_{-k}$ ($k>0$), $\deg a_{-k}=\deg b_{-k}=k$. Define two representations of this algebra $\pi$ and $\bpi$ on $\bcF_p$:
\eq$$
\Aligned{
\pi(a_{-k})
&={\alpha^-_k-\alpha^+_k\over A^{(1)}_k},
&\pi(b_{-k})
&={\beta^-_k-\beta^+_k\over A^{(2)}_k},
\\
\bpi(a_{-k})
&={\alpha^-_{-k}-\alpha^+_{-k}\over A^{(1)}_k},
&\bpi(b_{-k})
&={\beta^-_{-k}-\beta^+_{-k}\over A^{(2)}_k}.
}\label{piabdef}
$$
To any element $h\in\cA$ we may associate a state in $\cF_p$ and a state in~$\bcF_p$:
\eq$$
{}_p\langle h|={}_p\langle1|\pi(h),
\qquad
|h\rangle_p=\bpi(h)|1\rangle_p
\label{hvecsdef}
$$
For any pair of elements $h,h'\in\cA$ define the functions
\eq$$
J^{h\bh'}_{KL,p}(X|Y)
={}_p\langle h|t(x_1)\ldots t(x_K)\tdag(y_1)\ldots\tdag(y_L)|h'\rangle_p.
\label{Jhbhdef}
$$
These functions are straightforwardly calculated for any given pair $h,h'$. Indeed, move $\pi(h)$ to the right and $\bpi(h)$ to the left with the use of the commutation relations
\eq$$
\Aligned{{}
[\pi(a_{-k}),\lambda_\pm(z)]
&=(\mp)^{k+1}z^k\lambda_\pm(z),
&[\pi(a_{-k}),\lambdadag_\pm(z)]
&=(\mp)^{k+1}z^k\lambdadag_\pm(z),
\\
[\pi(b_{-k}),\lambda_\pm(z)]
&=(\mp)^kz^k\lambda_\pm(z),
&[\pi(b_{-k}),\lambdadag_\pm(z)]
&=-(\mp)^kz^k\lambdadag_\pm(z),
\\
[\lambda_\pm(z),\bpi(a_{-k})]
&=-(\mp)^{k+1}z^{-k}\lambda_\pm(z),
&[\lambdadag_\pm(z),\bpi(a_{-k})]
&=-(\mp)^{k+1}z^{-k}\lambdadag_\pm(z),
\\
[\lambda_\pm(z),\bpi(b_{-k})]
&=-(\mp)^kz^{-k}\lambda_\pm(z),
&[\lambdadag_\pm(z),\bpi(b_{-k})]
&=(\mp)^kz^{-k}\lambdadag_\pm(z)
}\label{piablambdacommut}
$$
and
\eq$$
\Gathered{{}
[\pi(a_{-k}),\bpi(a_{-k})]=-(1+(-1)^k)k(A^+_k)^{-1},
\qquad
[\pi(b_{-k}),\bpi(b_{-k})]=-(1-(-1)^k)k(B^+_k)^{-1},
\\
[\pi(a_{-k}),\bpi(b_{-k})]=[\pi(b_{-k}),\bpi(a_{-k})]=0.
}\label{piabcommut}
$$
At last, $\pi$ on the right and $\bpi$ on the left kill the vacuums:
\eq$$
\pi(a_{-k})|1\rangle_p=\pi(b_{-k})|1\rangle_p=0,
\qquad
{}_p\langle1|\bpi(a_{-k})={}_p\langle1|\bpi(b_{-k})=0.
\label{piab-vac}
$$

The main fact is that the functions
\eq$$
F^{h\bh'}_p(\theta_1,\ldots,\theta_K,\theta'_1,\ldots,\theta'_L)
_{\underbrace{\scriptstyle1\ldots1}_K\underbrace{\scriptstyle\bar1\ldots\bar1}_L}
=J^{h\bh'}_{KL,p}(\e^{\theta_1},\ldots,\e^{\theta_K}|\e^{\theta'_1},\ldots,\e^{\theta'_L})
\cR_p(\theta_1,\ldots,\theta_K|\theta'_1,\ldots,\theta'_L)
\label{fhbhp-def}
$$
again satisfy the form factor axioms and define operators $\Phi^{h\bh'}_{pq}$. In the same way as in~\cite{Feigin:2008hs} we conclude that they correspond to descendant operators over $\Phi_{pq}$. The operators $\Phi^h_{pq}$ are right (chiral) descendants, while the operators $\Phi^\bh_{pq}$ are left (antichiral) ones. If $h\in\cA_k$, $h'\in\cA_{\bar k}$ the spin and ultraviolet conformal dimension of the operator $\Phi^{h\bh'}_{pq}$ are given by~(\ref{sd-exponetial}) if we assume~(\ref{ab-pp}).

In simple cases descendant operators can be easily understood. For example, $\Phi^{a_{-1}}_{pq}=-2\i m^{-1}\d_-\Phi_{pq}$, $\Phi^{\ba_{-1}}_{pq}=2\i m^{-1}\d_+\Phi_{pq}$. This can be generalized. Let
\eq$$
\iota_{1-2k}=a_{1-2k},
\qquad
\iota_{-2k}=b_{-2k}.
\label{IMdef}
$$
The element $\iota_{-k}$ in the $\pi$ representation only produces the factor $\sum^K_{i=1}\e^{k\theta_i}-(-1)^k\sum^L_{i=1}\e^{k\theta'_i}$ in the form factors, which is an eigenvalue of an integral of motion. We conclude that
\eq$$
\Phi^{\iota_{-k}h\,\bh'}_{pq}(x)=[\Phi^{h\bh'}_{pq}(x),I_k],
\qquad
\Phi^{h\,\overline{\iota_{-k}h'}}_{pq}(x)=[\Phi^{h\bh'}_{pq}(x),I_{-k}],
\label{iota-I}
$$
where $I_k$ is an appropriately normalized spin $k$ conserved charge. In other cases the identification of operators defined by their form factors with the operators (\ref{descendant-def}) is not clear.

An important property of form factors of descendant operators is the cluster factorization. Suppose that the sets $X$ and $Y$ decompose in two parts: $X=(\e^\Lambda X^{(1)})\cup X^{(2)}$, $Y=(\e^\Lambda Y^{(1)})\cup Y^{(2)}$. Denote $K^{(1)}=\#X^{(1)},\ldots,L^{(2)}=\#Y^{(2)}$. Let $h\in\cA_k$, $h'\in\cA_l$. We want to find the function $J^{h\bh'}_{KL,p}(X|Y)$ in the limit $\Lambda\to\infty$. Since in this limit $f_\pm(x\e^{-\Lambda})\to\omega^{\mp1/2}$ and $g_\pm(x\e^{-\Lambda})\to\omega^{\pm1/2}$, we obtain
\Multline$$
\e^{-k\Lambda}J^{h\bar h'}_{KL,p}(\e^\Lambda X^{(1)},X^{(2)}|\e^\Lambda Y^{(1)},Y^{(2)})
\\
=J^h_{K^{(1)}L^{(1)},p-\sigma(K^{(2)}-L^{(2)})}(X^{(1)}|Y^{(1)})
J^{\bar h'}_{K^{(2)}L^{(2)},p+\sigma(K^{(1)}-L^{(1)})}(X^{(2)}|Y^{(2)})
+O(\e^{-\Lambda})\quad\text{as $\Lambda\to\infty$.}
\label{ClusterProp}
$$
It means that for large difference of two groups of rapidities each form factor of a $(k,\bar k)$ level operator factorizes into the product of form factors of a right $(k,0)$ level descendant operator over $V_{abq}(x)$ and a left $(0,\bar k)$ level one.

\section{Generating functions of descendant operators}
\label{sec-genfunc}

The currents $t(z),\tdag(z)$ together with $s_\pm(z)$ defined in~(\ref{ttdag-pole}), (\ref{s-def}) generate a current algebra. The currents $s_\pm(z)$, which have no analogue in the usual free field representation~\cite{Jimbo:2000ff,Fateev:2006js,Fateev:2008zz}, play a special role. Insertion of one of them into a matrix element ${}_p\langle h|\cdots|h'\rangle_p$ reduces to multiplying the matrix element by an $h,h'$-independent factor. This factor provides the correct residues of the kinematic poles in form factors. Besides, the currents $s_\pm(z)$ play an important role in the construction of generating functions of descendant operators below.

Consider the products
\eq$$
\Aligned{
T^{(\infty)}(z;X;Y;U;V)
&=\prod^{\#X}_{i=1}t\left(1\over zx_i\right)\prod^{\#Y}_{i=1}\tdag\left(1\over zy_i\right)
\prod^{\#U}_{i=1}s_+\left(1\over zu_i\right)\prod^{\#V}_{i=1}s_-\left(1\over zv_i\right),
\\
T^{(0)}(z;X;Y;U;V)
&=\prod^{\#X}_{i=1}t(zx_i)\prod^{\#Y}_{i=1}\tdag(zy_i)
\prod^{\#U}_{i=1}s_+(zu_i)\prod^{\#V}_{i=1}s_-(zv_i).
}\label{TXYUVdef}
$$
They make it possible to define infinite sets of vectors
\eq$$
\Aligned{
{}_p\langle1|T^{(\infty)}(z;X;Y;U;V)
&=\sum^\infty_{n=0}{}_p\langle n;X;Y;U;V|z^{-n},
\\
T^{(0)}(z;X;Y;U;V)|1\rangle_p
&=\sum^\infty_{n=0}z^n|n;X;Y;U;V\rangle_p.
}\label{vecexpansion}
$$
But these vectors do not belong to $\cF_p,\bcF_p$ since they contain factors $\omega^{r\gamma^++(\#X-\#Y-r)\gamma^-}$ with $r\in\Z$. Nevertheless, the contribution of these factors in each matrix element can be, in fact, reduced to a shift of the zero mode~$p$ and a constant factor.

Now we formulate the main statement about the generating functions for descendant operators. For any given numbers $p,n,n'$ and $L-K=q$, and sets $\bar X,\bar Y,\bar U,\bar V,\bar X',\bar Y',\bar U',\bar V'$, subject to the conditions%
\subeq{\label{XYUVconds}
\Align$$
&(-)^kS_k(\bar X)=S_k(\bar Y)=-S_k(\bar U)-S_k(\bar V)
&&(1\le k\le n),
\label{XYUVinfcond}
\\
&(-)^kS_k(\bar X')=S_k(\bar Y')=-S_k(\bar U')-S_k(\bar V')
&&(1\le k\le n'),
\label{XYUV0cond}
$$}
there exists a pair of elements $h\in\cA_n$, $h'\in\cA_{n'}$, which are the same for any value of $K$ (for given $q$), such that
\eq$$
J^{h\bh'}_{KL,p}(X|Y)={}_{p_0}\langle n;\bar X;\bar Y;\bar U;\bar V|t(x_1)\ldots t(x_K)
\tdag(y_1)\ldots\tdag(y_L)|n';\bar X';\bar Y';\bar U';\bar V'\rangle_{p_0}
\label{XYUVmatrel}
$$
with $p_0=p+(\#\bar X'-\#\bar Y'-\#\bar X+\#\bar Y)\sigma$. Moreover, for generic values of $p$ the whole spaces $\pi(\cA)\subset\cF_p$ and $\bpi(\cA)\subset\bcF_p$ can be spanned by such vectors. The proof of this statement repeats in main steps that of the analogous statement in~\cite{Feigin:2008hs} in the case of sinh-/sine-Gordon model. We postpone it to a more detailed future publication.

We may say that matrix elements between vectors in the l.h.s.\ of (\ref{vecexpansion}) are generating functions of form factors of descendant operators subject to the conditions~(\ref{XYUVconds}). Note that without presence of the currents $s_\pm(z)$ these conditions have no nontrivial solutions. Another fact to be noted is that for special values of $p$ the `spanning' property can be broken. It is broken e.g.\ for $p_\pm=g^{-1}+1$, $q=0$, which corresponds to the unit operator, whose form factors all vanish except the zero-particle one. It seems to be broken of all $(a,b)\in\Z^2$, $q+a+b\in2\Z$.

There is an important consequence of this statement. From (\ref{ppreflection}) we obtain
\Align*$$
&{}_{p_0}\langle n;\bar X;\bar Y;\bar U;\bar V|t(x_1)\ldots t(x_K)
\tdag(y_1)\ldots\tdag(y_L)|n';\bar X';\bar Y';\bar U';\bar V'\rangle_{p_0}
\\
&\qquad\qquad
={J_{-q_0,p_{0+}}\over J_{-q_0,-p_{0-}}}\cdot
{}_{-p_0^*}\langle n;\bar X;\bar Y;\bar U;\bar V|t(x_1)\ldots t(x_K)
\tdag(y_1)\ldots\tdag(y_L)|n';\bar X';\bar Y';\bar U';\bar V'\rangle_{-p_0^*},
$$
where $q_0=q-\#\bar X+\#\bar Y-\#\bar X'+\#\bar Y'$. The r.h.s.\ is equal to $J^{\tilde h\bar{\tilde h}'}_{KL,-p^*}(X|Y)$ with some elements $\tilde h,\tilde h'\in\cA$. We conclude that
for a given generic value of $p$ and any given value of $q$ for all $h\in\cA_n$, $h'\in\cA_{n'}$ there exist elements $\tilde h\in\cA_n$, $\tilde h'\in\cA_{n'}$ such that
\eq$$
J^{h\bh}_{KL,p}(X|Y)=J^{\tilde h\bar{\tilde h}'}_{KL,-p^*}(X|Y).
\label{ppreflection-descendants}
$$
This means that the descendant operators possess the reflection property, though it is not as trivial as that in the case of primary operators.

\section{Conclusion}

Here we found form factors of quasilocal operators in the complex sinh-Gordon model within the algebraic approach. The form factors of exponential operators are found completely up to overall normalization factors. For descendant operators, we have different definitions in terms of the basic field $\vartheta,\varphi$ and in terms of form factors. More or less direct identification is only accessible in special cases related to the action of integrals of motion. We study main properties of form factors and show how the reflection relations manifest themselves in the form factor theory.

The results can be also applied to the complex sine-Gordon theory ($g<0$), but its complicated spectrum of bound states demands more accuracy. In the special case of $g^{-1}$ negative integer the construction describes the $\Z_N$ Ising model, but in this case a selection of admissible operators is necessary. For primary fields this selection is well-known, but for descendant operators it is complicated. We will describe it in a forthcoming paper. There we will also give proofs omitted here for brevity.

\section*{Acknowledgments}

We are grateful A.~Litvinov for a useful discussion. The study of form factors of descendant operators was supported by Russian Science Foundation under the grant \#~14-12-01383.

\raggedright
\bibliographystyle{mybib}
\bibliography{main}

\end{document}